\documentclass[english]{lni}
\usepackage{psfrag}
\usepackage{booktabs}
\usepackage{graphicx, cite}
\usepackage{amsmath, amssymb}

\begin{document}
\title{Topic Extraction and Bundling of Related Scientific Articles}

\author{Shameem A Puthiya Parambath\\
       Umea University, Umea, Sweden.\\
       email: shameem.ahamed@gmail.com
}

\maketitle

\begin{abstract}
\begin{abstract} 
Automatic classification of scientific articles based on common characteristics is an interesting problem with many applications in digital library and information retrieval systems. Properly organized articles can be useful for automatic generation of taxonomies in scientific writings, textual summarization, efficient information retrieval etc. Generating article bundles from a large number of input articles, based on the associated features of the articles is tedious and computationally expensive task. In this report we propose an automatic two-step approach for topic extraction and bundling of related articles from a set of scientific articles in real-time. For topic extraction, we make use of Latent Dirichlet Allocation (LDA) topic modeling techniques and for bundling, we make use of hierarchical agglomerative clustering techniques. \\

We run experiments to validate our bundling semantics and compare it with existing models in use. We make use of an online crowdsourcing marketplace provided by Amazon called Amazon Mechanical Turk to carry out experiments. We explain our experimental setup and empirical results in detail and show that our method is advantageous over existing ones.      
\end{abstract}
\end{abstract}
    %
%\thispagestyle{plain}
    %
%--------------------------------------------------------
\section{Introduction}
With the advancement of information retrieval systems, especially search technologies, finding relevant information about any topic under the sky is relatively an easy task. Search engines like Google are very effective and popular for web retrieval. Researchers rely on these search engines to gather related works relevant to their field of work. Most of the search engines run dedicated services for scientific literature search, example includes popular websites like Google Scholar\cite{SCLR} and CiteSeerX\cite{CITE}. All of the websites above mentioned are very competent and retrieve large number of articles on proper input query. For example, our search for scholarly articles for the topic 'topic modeling' resulted in 1,190,000 and 141,843 articles using Google Scholar and CiteSeerX respectively. These results are ordered based on the indexing and ranking algorithms used by the underlying search system and contain similar articles scattered over different pages.\\

Grouping or bundling of articles, resulting from any extensive search into smaller coherent groups is an interesting but a difficult task. Even though lots of research studies were conducted in the area of data bundling, a concrete generalized algorithm does not exist. Effective grouping of data requires a precise definition of closeness between a pair of data items and the notion of closeness always depend on the data and the problem context. Closeness is defined in terms of similarity of the data pairs which in turn is measured in terms of dissimilarity or distance between pair of items. In this report we use the term similarity,dissimilarity and distance to denote the measure of closeness between data items. Most of the bundling scheme start with identifying the common attributes(metadata) of the data set, here scientific articles, and create bundling semantics based on the combination of these attributes. Here we suggest a two step algorithm to bundle scientific articles. In the first step we group articles based on the latent topics in the documents and in the second step we carry out agglomerative hierarchical clustering based on the inter-textual distance and co-authorship similarity between articles. We run experiments to validate the bundling semantics and to compare it with \emph{content only} based similarity. We used 19937 articles related to Computer Science from arviv \cite{ARXV} for our experiments.

\section{Topic Extraction}
\subsection{Latent Dirichlet Allocation}

Latent Dirichlet Allocation(LDA)\cite{DAM03} is a probabilistic generative model for document modeling. It is based on Probabilistic Latent Semantic Analysis(PLSA), a generative model suggested by Thomas Hofmann in \cite{Hofmann99,TH99}. \cite{DAM03} LDA is based on dimensionality reduction assumption, \emph{bag-of-words} assumption i.e., order of words in a document is not important. Words are considered to be conditionally independent and identically distributed. Ordering of documents is also neglected and assumed to be independent and identically distributed. This is called document exchangeability. Same principle applies for topics also. There is no prior ordering of topics which makes it identifiable. Basic assumption in the LDA model are given below
\begin{itemize}
\item Number of documents are fixed
\item Vocabulary size is fixed
\item Number of topics are fixed
\item Word distribution is a multinomial distribution
\item Topic distribution is a multinomial distribution
\item topic weight distribution is a Dirichlet distribution
\item word distribution per topic is a Dirichlet distribution
\end{itemize} 
Generative model suggest a probabilistic procedure to generate documents given a distribution over topics. Given a distribution over topics, a document can be generated by recursively selecting a topic over given topic distribution and then selecting a word from the selected topic. We, now, formally define the mathematical model behind LDA. We are given a fixed set of documents $D=\{d_1,d_2,d_3,....d_N\}$, a fixed set of vocabulary $W=\{w_1,w_2,w_3,....w_M\}$ and a set of topics $T=\{t_1,t_2,...,t_k\}$. Let $d \in D$ denote a random document,$w \in W$ denote a random word and $t \in T$ denote a random topic. Let $P(d)$ be the probability of selecting a document, $P(t|d)$ is the probability of selecting topic $t$ in document $d$ given the probability of selecting the document $d$ and $P(w|t)$ is the probability of selecting word $w$ in topic $t$ given the probability of selecting the topic $t$. The join probability distribution of the observed variables $(d,w)$ is
%\begin{center}
\begin{align*}
 & P(d,w) = P(d)P(w|d)\\
 &\textit{Since w \& d are conditionally independant over t}\\
 & P(w|d) = \sum_{t_1}^{t_k} P(w|t)P(t|d) \implies P(d,w) = P(d)\sum_{t_1}^{t_k} P(w|t)P(t|d)\\
 & \textit{According to Baye's Rule}\\
 & P(d)P(t|d) = P(d|t)P(t)  \implies P(d,w) = \sum_{t_1}^{t_k} P(t)P(w|t)P(d|t) 
\end{align*}
%\end{center}
Now thinking the opposite direction, given a document, using statistical inference we can find the topics associated with each document. This illustrates the statistical inference problem, inverse of the approach mentioned above. Here given a document, we would like to find the associated topics which is most likely to have generated the given document. We refer to the set of topics generated using topic modeling method as topic classes. This involves inferring the word distribution in the topics and topic distribution in the documents given the word distribution in the documents. LDA algorithm generates this topic classes using statistical inference techniques based on the assumptions given earlier. LDA tool we used, MALLET, uses an algorithm based on Gibbs sampling\cite{CG92} to estimate the topic classes.

\section{Bundling}
In this section, we will elaborate the second step of our algorithm i.e., bundling documents in a given topic class. A topic, selected from the set of topics generated by LDA, is given to the clustering system and it generates coherent bundles based on the selected similarity measures. Similarity measures for our data set is defined based on extended co-authorship and inter-textual distance.

\subsection{Extended Co-authorship Dissimilarity}
Extended Co-authorship Dissimilarity between two articles is conceived in terms of the similarity between the extended co-authors of the articles. Extended co-authors is defined as the union of the set of authors and referenced authors of an article. Extended Co-authorship Similarity between two articles is defined as the Jaccard Coefficient on the extended co-authors two articles.
\begin{align*}
	SIM(A,B)=\frac{|Extended\ Co-auth(A) \cap\ Extended\ Co-auth(B)|}{|Extended\ Co-auth(A)\ \cup\ Extended\ Co-auth(B)|},\ where &\\
    Extended\ Co-auth(A) =\ Extended\ Co-authorship\ of\ article\ A &\\
    Extednded\ Co-auth(B) =\ Extended\ Co-authorship\ of\ article\ B &
\end{align*}
Corresponding Extended Co-authorship Dissimilarity is defined as $ExtCoauth(A,B)\ =\ 1 - SIM(A,B)$. We create a proximity matrix $\textbf{ExtCoauth}$ containing the Extended Co-authorship Dissimilarity among all the articles as $\textbf{ExtCoauth} = [ExtCoauth_{i,j}]_{n*n} = [ExtCoauth(i,j)]$

\subsection{Inter-textual Distance}
Inter-textual distance, due to Labbe\cite{LAB01}, is defined over the frequency of the common vocabulary of the texts. It measures the relative distance of the texts from each other. Mathematically inter-textual distance between two texts A and B is defined as,
\begin{align*} 
D_{(A,B)} =& \frac{\sum\limits_{V_A,V_{A(B)}}\|F_{iA} - E_{iA(B)}\|}{N_A + N_{A(B)}}
\end{align*}
Here, $F_{iA}$ is the frequency of the $i^{th}$ vocabulary in document A, $F_{iB}$ is the frequency of the $i^{th}$ vocabulary in document B, $E_{iA(B)}$ is the frequency of the $i^{th}$ vocabulary in B with mathematical expectation more than or equal to one with respect to A. $N_A$ is the sum of the frequency of vocabulary in A, $N_B$ is the sum of the frequency of vocabulary in B and $N_{A(B)}$ is the sum of frequency of vocabulary in B with expectation value more than or equal to one. A proximity matrix $\textbf{Cont}$ is constructed for all the articles in the given topic class containing the inter-textual distances between them.
\begin{align*} 
N_A = \sum\limits_{V_A}F_{iA},\ N_{A(B)} = \sum\limits_{V_B}E_{iA(B)}  \\
E_{iA(B)} = F_{iB} \times \frac{N_A}{N_B},\ N_B = \sum\limits_{V_B}F_{iB} 
\end{align*}
   
\subsection{Bundling}
To apply hierarchical, agglomerative clustering algorithm, we create a combined proximity matrix $\textbf{D}$ from the respective distance measures $\textbf{ExtCoauth}$ and $\textbf{Cont}$ as given by 
\begin{align*}
\textbf{D} =& \alpha * \textbf{ExtCoauth} + (1-\alpha) * \textbf{Cont}, 0 \leq \alpha \leq 1
\end{align*} 
 where $\alpha$ is the weight factor. We apply fastcluster algorithm as in \cite{DANM11} for hierarchical agglomerative clustering to create $\sqrt{n}$ number of bundles, where $n$ is the number of articles in the selected topic class.

\section{Experiments}
In this section, we detail the experimental protocol based on Amazon Mechanical Turk, a crowdsourcing market place of Amazon. There are two types of evaluation techniques employed to measure the quality of clustering schemes,  one being theoretical evaluation and other being user evaluation. We make use of user evaluation techniques here.\\
Amazon Mechanical Turk(AMT)\cite{AMT} is a crowdsourcing marketplace service provided by Amazon where users can work on small tasks which is currently difficult to achieve using computers i.e., work that requires human intelligence. In Mechanical Turk terminologies, tasks are called Human Intelligence Tasks(HIT), user who provides task is called Requester and user who works on the task is called Worker. A HIT is a well explained, self-contained question of the type described earlier. A requester will create HITs and publish it on AMT. A requester can assert some mechanism to recruit suitable workers or filter out unskilled workers through qualification tests. To run the experiments, we selected three topic classes from the set of twenty six topic classes. Topic classes selected for the experiments are Machine Learning, Information Retrieval and Graph Theory. We selected five bundles from these three topic classes. Each of these bundle is presented to users to check the quality and compare it with bundles generated using content based only clustering.

\subsection{Independent Study}
In independent study, we measure the quality of the bundling process independently through Worker feedback. Here we validate the semantics by asking the Worker to comment on the quality of the bundles generated by our algorithm. Here we make use of survey questionnaire in which we ask the Workers are asked to read the articles in the bundle and give their feedback on the similarity of the articles in the bundle.

\subsubsection{Results}
%\paragraph{Results}
Results of the survey questions are detailed in Table 1. All the 29 users participated in the survey for topic information retrieval confirm that the articles in the bundles are very similar which is 100\% success ratio. Out of the 24 users who participated in the survey for Graph Theory 20 users affirm that the member articles in the bundles are similar. In case of Machine Learning, 84.2\% of the participated workers agree with the member article similarity in the bundle. Overall the agreement ratio of independent study is 89.1\% which is a very good indication that our selection of clustering semantic is very good. 

\begin{center}
\begin{table}[ht]
%\begin{tabular}{|p{5cm}|p{3cm}|p{3cm}|p{4cm}|}
\centering
\begin{tabular}{|c|c|c|c|}
\hline
\hline \textbf{Topic} & \textbf{Agreement} & \textbf{Non-agreement} & \textbf{Agreement Ratio} \\ \hline
Information Retrieval & 29 & 0 & 100\% \\ \hline
Graph Theory & 20 & 4 & 83.3\% \\ \hline
Machine Learning & 32 & 6 & 84.2\% \\ \hline
Overall & - & - & 89.1\% \\ \hline
\end{tabular}
\caption{Independent Study Results}
\end{table}
\end{center}

\subsection{Comparative Study}
In comparative study, we ask the worker to do "side-by-side" comparison of two bundling results one based on the content and extended authorship similarity and the other one based on content similarity only. Aim of the comparative study is to check whether the semantic used  by us gives a better result than the other popular commonly used semantics. We employ survey type questionnaire in which we ask the worker to read the two bundles and give each bundle most appropriate name. At the end they are asked to point the bundle, which was easiest to name. Our assumption is diverse bundle will be difficult to name and similar bundle will be very easy to name.

\subsubsection{Results}
Results of the comparative study is given in the tables 2,3, 4 and 5. Table 5.2 contains the result of the survey questionnaire for the topic information retrieval. Overall a total of 50 users per topic class participated in the evaluation. 82.7\% of the workers selected the extended co-authorship + content based similarity over only content based similarity.

\begin{center}
\begin{table}[ht]
%\begin{minipage}[b]{0.45\linewidth}
\centering
\begin{tabular}{|c|c|c|}
\hline 
\hline \textbf{Batch} & \textbf{Preferred} & \textbf{Not Preferred}\\ \hline
1 & 8 & 2 \\ \hline
2 & 8 & 2 \\ \hline
3 & 10 & 0 \\ \hline
4 & 8 & 2 \\ \hline
5 & 9 & 1 \\ \hline
Result & 86\% & 14\% \\ \hline
\end{tabular}
\caption{Comparative Study Results Information Retrieval}
%\end{minipage}
%\hspace{0.5cm}
\end{table}
\end{center}

\begin{center}
\begin{table}[ht]
%\begin{minipage}[b]{0.45\linewidth}
\centering
\begin{tabular}{|c|c|c|}
\hline
\hline \textbf{Batch} & \textbf{Preferred} & \textbf{Not Preferred}\\ \hline
1 & 8 & 2 \\ \hline
2 & 8 & 2 \\ \hline
3 & 9 & 1 \\ \hline
4 & 7 & 3 \\ \hline
5 & 9 & 1 \\ \hline
Result & 82\% & 18\% \\ \hline
\end{tabular}
\caption{Comparative Study Results Graph Theory}
%\end{minipage}
%\hspace{0.5cm}
\end{table}
\end{center}

%\begin{minipage}[b]{0.45\linewidth}\centering
\begin{center}
\begin{table}[ht]
\centering
\begin{tabular}{|c|c|c|}
\hline
\hline \textbf{Batch} & \textbf{Preferred} & \textbf{Not Preferred}\\ \hline
1 & 7 & 3 \\ \hline
2 & 9 & 1 \\ \hline
3 & 9 & 1 \\ \hline
4 & 8 & 2 \\ \hline
5 & 7 & 3 \\ \hline
Result & 80\% & 20\% \\ \hline
\end{tabular}
\caption{Comparative Study Results Machine Learning}
%\end{minipage}
\end{table}
\end{center}

%\begin{minipage}[b]{0.45\linewidth}\centering
\begin{center}
\begin{table}[!h]
\centering
\begin{tabular}{|c|c|c|}
\hline 
\hline & \textbf{Preference Ratio} & \textbf{Non-Preference Ratio}\\ \hline
Result & 82.7\% & 17.3\% \\ \hline
\end{tabular}
\caption{Comparative Study Results (Combined)}
%\end{minipage}
\end{table}
\end{center}

\section{Conclusion}
Our algorithm gives very promising result when used with unstructured data and we believe that with structured data it will yield far better results. User study conducted on the unstructured data set shows very positive indication towards the effectiveness of our bundling semantics. Our algorithm can be easily extended by using other similarity measures like Year-of-Publishing, co-authorship graphs, keywords etc. 

\bibliography{ref}

\end{document}